\documentclass[reprint,
superscriptaddress,
nofootinbib,
 amsmath,amssymb,
 aps,
prd,
floatfix,
]{revtex4-2}
\usepackage{graphicx}
\usepackage{bm}
\usepackage{epsfig}
\usepackage{amsmath}
\usepackage{hyperref}
\usepackage{natbib}
\usepackage{color}

\newcommand{\rh}{\ensuremath{r_{1/2}}}
\newcommand{\Rh}{\ensuremath{R_{1/2}}}
\newcommand{\vh}{\ensuremath{v_{1/2}}}
\newcommand{\mfdm}{\ensuremath{m_{\rm FDM}}}

\begin{document}

\title{Not so fuzzy: excluding FDM with sizes and stellar kinematics\\ of ultra-faint dwarf galaxies}
\author{Neal Dalal}
\email{ndalal@pitp.ca}
\affiliation{Perimeter Institute for Theoretical Physics, 31 Caroline Street N., Waterloo, Ontario, N2L 2Y5, Canada}
\author{Andrey Kravtsov}
\affiliation{Department of Astronomy \& Astrophysics, The University of Chicago, Chicago, IL 60637 USA}
\affiliation{Kavli Institute for Cosmological Physics, The University of Chicago, Chicago, IL 60637 USA}
\affiliation{Enrico Fermi Institute, The University of Chicago, Chicago, IL 60637 USA}

\begin{abstract}
We use observations of ultra-faint dwarf (UFD) galaxies to constrain the particle mass of ultra-light  dark matter. Potential fluctuations created by wave interference in virialized ``fuzzy'' dark matter (FDM) halos dynamically heat stellar orbits in UFDs, some of which exhibit velocity dispersions of $\lesssim 3$ km/s and sizes $\lesssim 40$ pc.  Using simulations of FDM halos, and existing measurements of sizes and stellar radial velocities in Segue 1 and Segue 2 UFDs, we derive a lower limit on the dark matter particle mass of $\mfdm > 3\times 10^{-19}$ eV at $99\%$ confidence, marginalized over host halo circular velocity. This constraint is conservative as it is derived under the assumption that soliton heating is negligible, and that no other sources of non-FDM dynamical heating of stars operate to increase velocity dispersion.
It can potentially be strengthened by future spectroscopic observations of additional stars in ultra-faint galaxies and by tightening theoretical constraints on the soliton size-halo mass relation. However, even the current conservative lower limit on the FDM mass makes this model indistinguishable from Cold Dark Matter at the scales probed by  existing astronomical observations. 
\end{abstract}

\maketitle

\section{Introduction}
\label{sec:intro}
The cold dark matter (CDM) model successfully describes a wide range of observations on scales larger than galaxies, but on sub-galactic scales,  deviations from CDM predictions remain viable.  One particularly interesting class of dark matter (DM) scenarios involves ultra-light particle masses, $m\sim 10^{-22}-10^{-20}$ eV, called fuzzy dark matter (FDM) \citep{Hu:2000ke,Hui:2016ltb,Hui2021,Ferreira2021}.  In this regime, the de Broglie wavelength of DM particles, $\lambda = h/mv$, can be large enough to produce observable wave effects on galactic scales \citep{Schive:2014dra}.  For example, a mass of $m=10^{-22}$ eV and a DM velocity dispersion of $v=200$ km/s would give $\lambda \sim 600$ pc. 

FDM models can nontrivially modify the small-scale linear power spectrum compared to CDM, leading to either suppression or enhancement of small-scale structure \cite{Hu:2000ke,Arvanitaki2020}, which in principle can be used to constrain FDM parameters in a manner similar to constraints on warm dark matter from the halo mass function (e.g., \cite{Nadler2021}).   
FDM halos also exhibit unique signatures distinct from warm dark matter halos. One example is the formation of dense solitons at the centers of FDM halos \citep{Schive:2014hza}, which can lead to core-like density profiles, rather than cuspy behavior.  

Early work using  FDM simulations found a tight scaling relation between the soliton mass and the virial mass of the host halo \cite{Schive:2014hza}, leading to a number of papers attempting to constrain the FDM particle mass by finding evidence for either the presence or absence of massive solitons in nearby galaxies \cite{Bar2018,Safarzadeh2020,Hayashi2021,Bar2021,Pozo2021}.  
More recent FDM simulations covering larger cosmological volumes, however, discount the existence of a tight scaling relation between soliton mass and host halo mass, and instead  indicate a large scatter in the correlation  \cite{May2021,Chan2022}.  Recent work has shown that FDM solitons can stably persist with mass either greatly exceeding \cite{Chan2022} or far below \cite{Yavetz2021} the original scaling relation \citep{Schive:2014hza}.  This casts doubt on whether reliable bounds on the FDM mass can be derived using the soliton-host scaling relation in the small number of nearby galaxies found in the local volume.  Note that this does not preclude some constraints to be placed on FDM masses using other scaling relations \cite{Zoutendijk2021a}.  

Another distinctive signature of FDM is that 
wave interference effects produce ubiquitous fluctuations in the local density and gravitational potential throughout FDM halos with contrast of order unity,  $\delta\rho\sim\rho$, and coherence lengths $r\approx \lambda/2\pi = \hbar/mv$ \cite{Schive:2014dra,Hui2021}. These fluctuations resemble transient granules of mass $\delta M \propto \rho/v^3$  that can gravitationally heat stellar systems (e.g., \citep{Hui:2016ltb,Church2019,Bar-Or2019,Bar-Or2021,Dalal2021}).  

This prediction of density and potential fluctuations from wave interference is robust, since it relies on simple, well-understood physics, namely the relation between the de Broglie wavelength and momentum: $\lambda=h/mv$.  This effect is therefore independent of other DM properties besides mass, like self-interactions.  
This stands in contrast to earlier FDM probes relying on poorly understood soliton scaling relations.  Unlike the scatter found in  soliton masses, there is no scatter in Planck's constant among different galaxies.  This means that a {\em single} galaxy can be used to derive tight constraints on the FDM particle mass, if that galaxy can be shown to be inconsistent with the effects of dynamical heating expected for light FDM masses. 

It is easy to see what galaxies would be most useful for constraining FDM heating in this way, using the following argument. The total enclosed halo mass at some radius $R$ is approximately $M\propto \rho\,R^3$, with a corresponding gravitational acceleration $a=GM/R^2 \propto \rho\,R$,
while in comparison, the gravitational acceleration caused by a mass perturbation $\delta M \propto \rho/v^3$ at distance $\lambda \propto v^{-1}$ is $\delta a = G \delta M/\lambda^2 \propto \rho/v$, giving a  fractional fluctuation in the gravitational acceleration  $\delta a/a \propto R^{-1} v^{-1}$. The heating effects of the fluctuations should therefore be strongest at the smallest radii $R$ and for the halos with the smallest internal velocities $v$.

At very small radii, the solitons discussed above can produce additional, strong dynamical heating.  Fluctuations in the soliton's size, amplitude, and centroid location, due to wave interference effects \citep{Li2021,Zagorac:2021}, can dominate the heating of central stars compared to the more diffusive effect described above \cite{Dutta2021}.
Stellar self-gravity can suppress this effect  \cite{Marsh2019,Chiang2021}, but in systems where gravity is completely dominated by dark matter, FDM heating can be extremely efficient.

These considerations point to the ultra-faint dwarf (UFD) galaxies \citep{Simon2019} as optimal systems to probe such heating.  The UFDs are DM-dominated, with mass to light ratios of $\gtrsim 300$ inside the 3D half-light radius $\rh$, which can be as small as $\rh \sim 50\,{\rm pc}\ll r_{\rm vir}$ in some of these galaxies.  The observed stellar velocity dispersions in these systems are correspondingly small, with $\sigma_v \approx 2-5$ km/s in many systems \citep{Simon2019}, while their stellar populations are predominantly very old \citep{Brown2012,Brown2014,Weisz2014}. 

Two examples of such systems are the nearby ultra-faint dwarfs Segue 1 \cite{Simon2011} and Segue 2 \cite{Kirby2013}, which are among the smallest and darkest galaxies known.  The projected (2D) half light radii of Segue 1 and Segue 2  are $\Rh=24.2 \pm 2.8$ pc and $\Rh=40.5\pm 3.0$ pc, respectively. Here we use the Plummer model $\Rh$ estimates of \cite{munoz_etal18}, with the values for Segue 2 corrected by a factor of 1.057 to account for the updated RR Lyrae-based distances of $37$ kpc to Segue 2 \cite{Nagarajan2021}, compared to the distance of 35 kpc used in \cite{munoz_etal18}.    Spectroscopic surveys of stars in these galaxies indicate extremely low velocity dispersions, with $\sigma_v = 3.9\pm 0.8$ km/s in Segue 1 \cite{Simon2011}, and only an upper limit $\sigma_v < 2.6$ km/s (95\% confidence) in Segue 2 \cite{Kirby2013}.

The stellar populations in UFDs are generically quite old, with ages of $\gtrsim 10$ Gyr \cite{Weisz2014}. For Segue 1, population modelling indicates that most stars were formed at $z>7$ \cite{Frebel2014,Webster2016}, while  modeling of Segue 2 \cite{Kirby2013} suggests a star formation history similar to that of the dwarf Ursa Minor, with typical age of $\gtrsim 9-10$ Gyr \cite{Carrera2002,Weisz2014}.
This implies that these stellar populations should have been exposed to the FDM-induced dynamical heating for correspondingly long ($\sim 10$ Gyr) time.  

We can approximately estimate the sensitivity of UFD galaxies to FDM heating using the following argument.  Let us approximate FDM fluctuations as granules of density fluctuation $\delta\rho \sim \rho$ and size $r=\lambda/2\pi = \hbar/(m \sigma_{\rm dm})$, where $\rho$ is the local mass density, and $\sigma_{\rm dm}$ is the velocity dispersion of dark matter, which may be different than the stellar velocity dispersion $\sigma_\star$, since the DM halo size is generally larger than the galaxy size.  Since $\delta\rho\sim\rho$, then the mass perturbation associated with a granule at location $R$ is $\delta M \approx (r/R)^3\,M$, where $M\approx (4\pi/3)\,\rho\,R^3$ is the enclosed mass at radius $R$, and we will consider locations of order the galaxy size, $R\approx \rh$, where $\rh$ is the 3D half-light radius.

In equilibrium, the stellar velocity dispersion $\sigma_\star$ is related to the galaxy size and enclosed mass by $M\approx 3 \sigma_\star^2 \rh/G$ \cite{wolf_etal10}, so the gravitational potential perturbation from this granule is $\delta\Phi\approx G\delta M/r \approx 3\, \sigma_\star^2 (r/\rh)^2$.  A granule passing by a star at a relative velocity of $\sim \sigma_{\rm dm}$ will perturb the star's velocity by $\delta v \sim \delta\Phi/\sigma_{\rm dm}$, and over a duration of time $t$, the star will encounter $N\sim \sigma_{\rm dm} t/r$ such granules.  Therefore, after time $t$ the variance in stellar velocities should increase by an amount $\Delta\sigma_\star^2 \approx N\,\delta v^2$, and using $r=\hbar/(m\sigma_{\rm dm})$ gives
\begin{equation}
\Delta \sigma_\star^2 \simeq 9 \left(\frac{\sigma_\star}{\sigma_{\rm dm}}\right)^4
\left(\frac{\hbar}{m}\right)^3 \frac{t}{\rh^4}.
\label{eqn:ballpark}
\end{equation}
We can therefore solve for the FDM mass $m$ that will cause the stellar velocity variance to double, $\Delta\sigma_\star^2\approx\sigma_\star^2$, after time $t$. Plugging in $t=10$ Gyr, $\rh=50$ pc, $\sigma_\star=3$ km/s, and $\sigma_{\rm dm} = 6$ km/s gives a mass $m\sim 10^{-19}$ eV.  This means that UFDs like Segue 1 and Segue 2 should be sensitive to the entire range of FDM masses, $m\sim 10^{-22}-10^{-20}$ eV, and potentially could detect or exclude this model.

Motivated by this estimate, in this study we use kinematic and structural measurements of Segue 1 and Segue 2 galaxies to derive constraints on the FDM particle mass.  We simulate the motion of stars moving in the FDM halos and compare to measurements of these two UFDs.  We will see that the above estimate is confirmed by our detailed numerical simulations, allowing us to place stringent bounds on ultra-light particle dark matter.

\section{Simulations}
\label{sec:sims}

\begin{figure*}
    \centering
    \includegraphics[width=0.48\textwidth,trim=1in 3in 1in 3in]{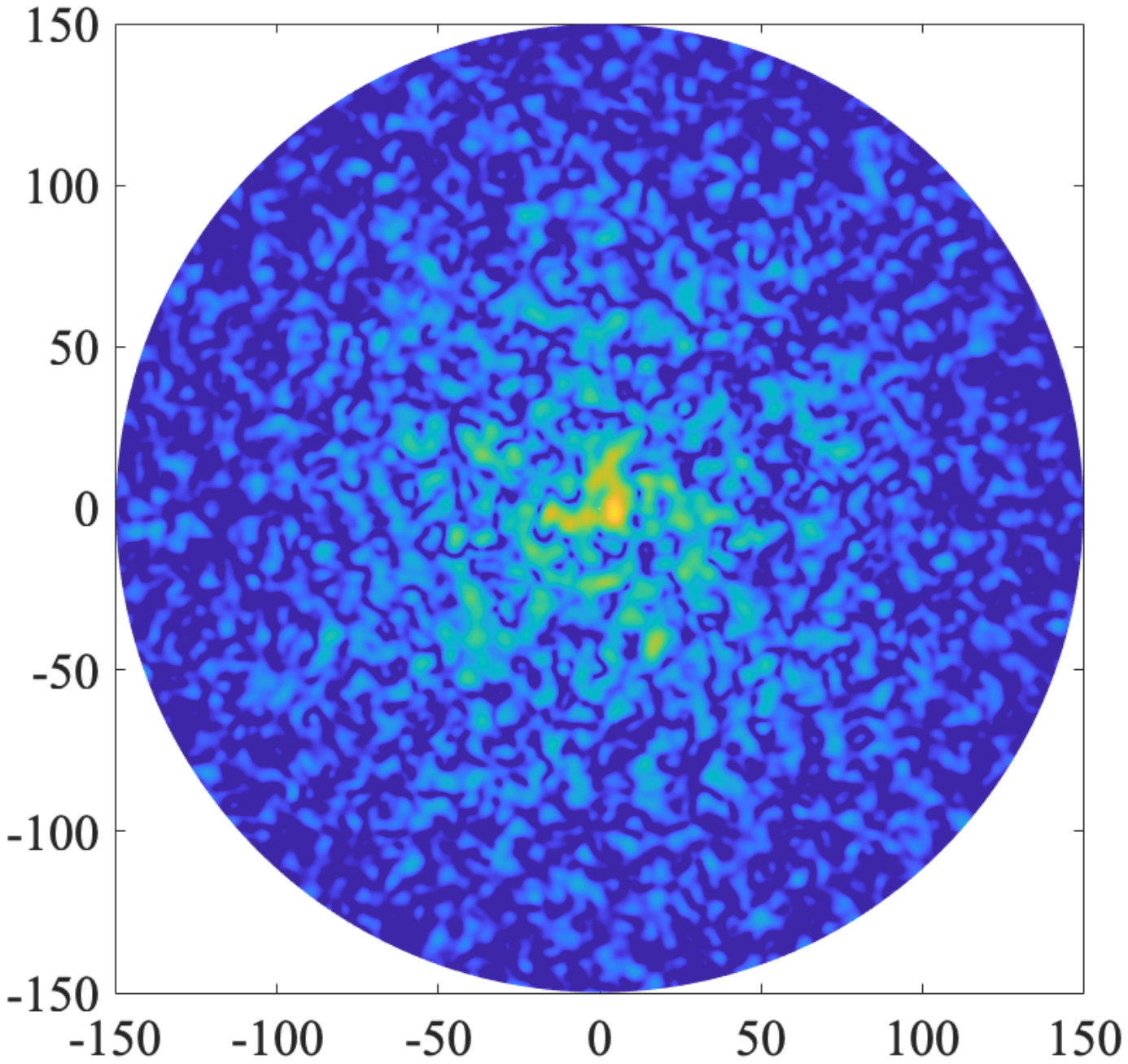}
    \includegraphics[width=0.48\textwidth,trim=1in 3in 1in 3in]{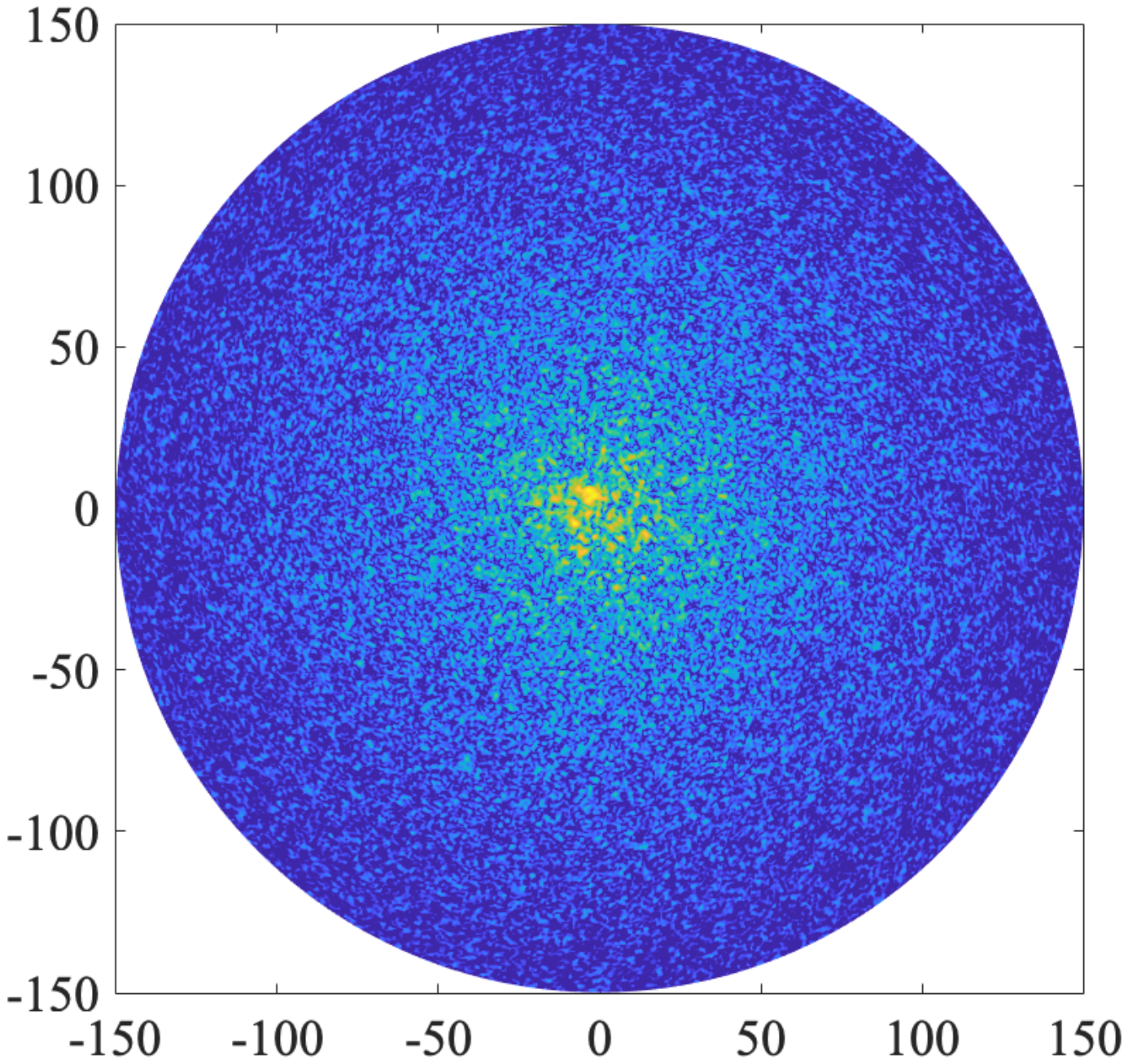}
    \caption{Example FDM density snapshots for NFW halos.  The left panel shows a simulation with $\mfdm=2\times10^{-19}$ eV, while the right panel has $\mfdm=8\times10^{-19}$ eV. Length units are parsecs, and the color scale is logarithmic in density $\rho$.  For comparison, the galaxies that we analyze have half-light radii $\Rh \simeq 25-40$ pc.}
    \label{fig:sim}
\end{figure*}

We simulate the motions of test stellar particles that are evolved in FDM halos, using an approximate perturbative treatment of the effects of FDM wave interference fluctuations presented in Ref.\ \citep{Dalal2021}.  
Briefly, we construct the FDM wavefunction as a sum over eigenfunctions of the Hamiltonian for a static gravitational potential; Fig.\ \ref{fig:sim} shows examples.  This approach is accurate to first order in perturbation theory and neglects the nonlinear interactions between different modes, which should be sufficient for our calculations, although it may be worthwhile to study whether greater accuracy can be achieved by going to higher order in time-dependent perturbation theory \cite{Zagorac:2021}.

We describe the mean halo potential using the NFW profile \cite{NFW1996}, however we have explicitly checked that our results are insensitive to the exact form of the halo profile (e.g., NFW, cored isothermal, etc.), and depend principally on the mass enclosed within the stellar half-light radius $\rh$ \cite{wolf_etal10}.  
The NFW profile is described by two independent parameters, such as virial mass and concentration ($M_{\rm vir}, c_{\rm vir}$), or central density and scale radius ($\rho_s, r_s$), and we are free to use any pair of independent parameters.  Since the results depend mainly on the parameter combination that holds fixed $\vh\equiv \left[GM(\rh)/\rh\right]^{1/2}$, we adopt $\vh$ as one of the parameters, and choose the maximum circular velocity $v_{\rm max}$ as the other independent NFW parameter.  Modelling of the abundance of  Milky Way dwarf satellites indicates that galaxies with ultra-faint luminosity are typically hosted by halos of mass $M_{200}\in 10^8-10^9\, M_\odot$ or maximum circular velocity of $v_{\rm max}\in 10-25\,\rm km/s$ \citep{Nadler2020,Kravtsov2021}, but in the analysis described below, we do not restrict the host halo parameters to this range.

FDM simulations have found that halo profiles are well described by the NFW profile, except for the presence of a central soliton \citep{Schive:2014hza}.
In our simulations, we can account for the central soliton by adding a soliton to the halo density, with a profile following the functional form found in Schr\"odinger-Poisson simulations \cite{Chiang2021}. This gives time-averaged density profiles that agree reasonably well with the desired input profile.  The density fluctuations in the soliton region resemble those found in full Schr\"odinger-Poisson simulations \cite{Chiang2021}, in agreement with previous work that found that soliton fluctuations can be explained by wave interference between the lowest eigenmodes \cite{Li2021,Zagorac:2021}.  Our simulations find that including the soliton can enormously enhance heating of stellar orbits, depending on the soliton mass, in agreement with previous work \cite{Dutta2021}.

Uncertainty in the expected sizes and masses of solitons  will complicate any attempt to constrain the FDM particle mass using soliton effects, as noted above.  Due to extremely efficient heating from the soliton, the uncertainty in soliton mass leads to uncertainty in the expected heating rates.  We therefore adopt a conservative approach, and compute constraints on the FDM mass using simulations without solitons.  This significantly weakens our derived constraints, and below we use simulations including solitons to indicate how much more strongly the FDM mass can be bounded if the theory uncertainty on the core-halo mass scaling can be reduced.  

We evolve test particles in these FDM potentials for 10 Gyr.  The test particles are initialized in a spherically symmetric distribution with the projected surface density profiles that decrease exponentially in radius consistent with a typical surface density profile in observed dwarf spheroidal galaxies of this luminosity \citep[e.g.,][]{Okamoto2012,mutlu_pakdil_etal18,munoz_etal18}. Such profile corresponds to the deprojected 3D density profiles $\propto K_0(r/r_0)$, where $r_0$ is the exponential scale length, and $K_0$ is a modified irregular Bessel function.  The initial velocities of the test particles are locally Maxwellian distributed, with a variance given by the solution to the static isotropic Jeans equation \cite{BinneyTremaine}.  Different choices for $r_0$ give different initial velocity dispersions, so we treat $r_0$ as a nuisance parameter and marginalize over it as a means of marginalizing over uncertain effects of star formation and baryonic heating.  

We initialize using isotropic velocity distributions, and this neglect of velocity anisotropy makes our derived constraints even more conservative, for the following reason.  In models with low \mfdm, FDM scattering rapidly isotropizes the velocity distribution, making any initial anisotropy irrelevant.  In contrast, for smooth potentials, or those with large \mfdm, heating is less efficient, and so the initial anisotropy can be preserved.  By neglecting anisotropy, we therefore are underestimating the parameter space over which smooth potentials can fit the observed data, in comparison to models with low \mfdm. This artificially lowers the likelihood for large \mfdm, making our constraints even more conservative. 

In summary, we simulate stars as test particles moving in the FDM halos for 10 Gyr using the method of Ref.\ \cite{Dalal2021}, with mean density profiles that follow the NFW profile with a given normalization treated as a nuisance parameter, and with stars initialized using isotropic Maxwellian velocity distributions that satisfy the Jeans equation.  We derive bounds on the FDM mass by marginalizing over the nuisance parameters: the halo mass (\vh, circular velocity within the stellar half-light radius) and the stellar size parameter $r_0$, using a suite of simulations that span the parameter space relevant for the Segue 1 and Segue 2 galaxies used in our constraint.  

\begin{figure}
    \centering
    \includegraphics[width=0.47\textwidth]{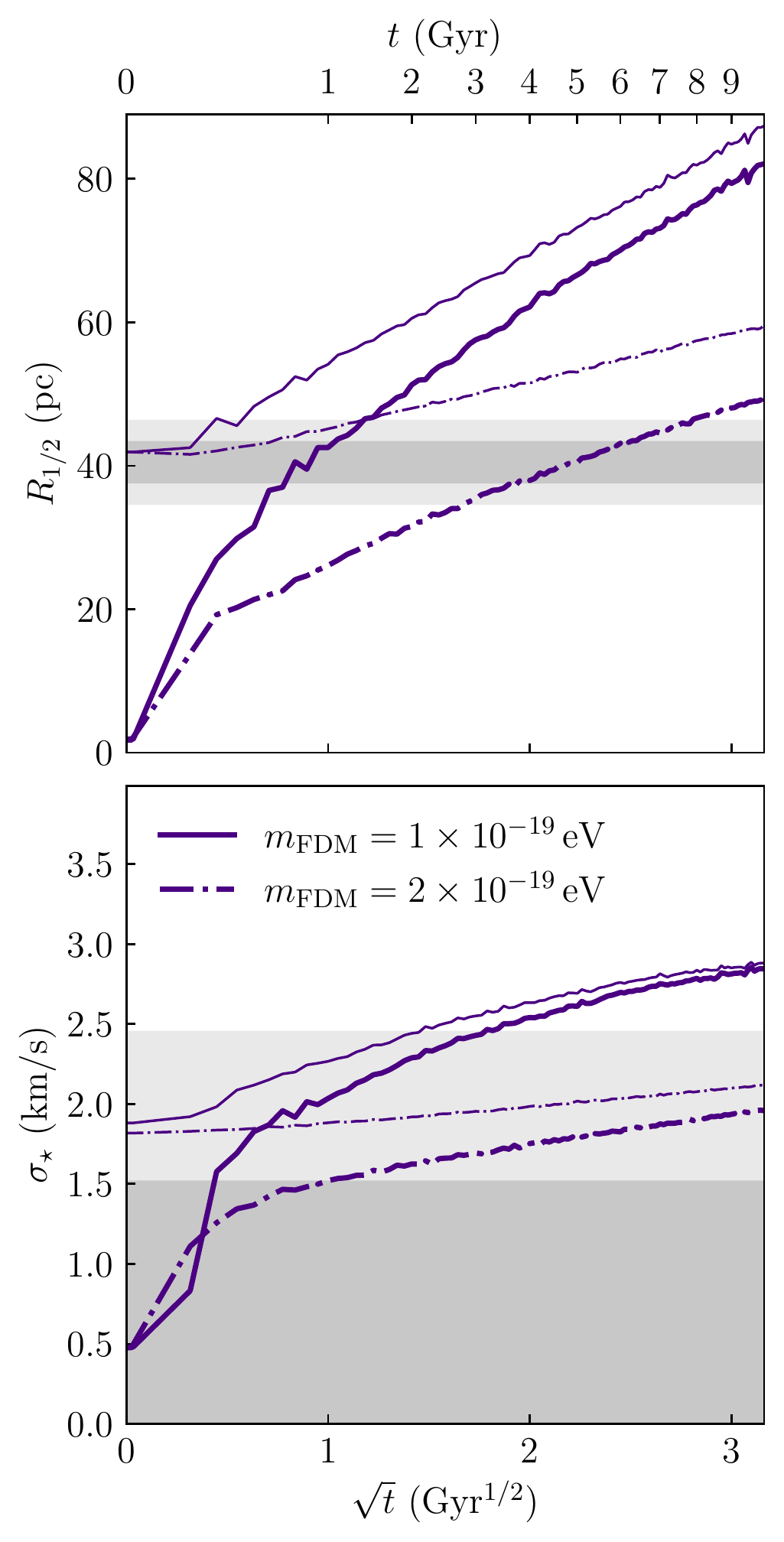}
    \caption{Evolution of galaxy properties in response to FDM heating.  In both panels, the 4 curves correspond to two different FDM masses and two different initial $r_0$, for host halo with $\vh=3$ km/s.  The upper panel shows growth in the projected half-light radius over time, while the bottom panel shows the growth in the velocity dispersion for stars within 3D radius $r=100$ pc.  For comparison, the grey bands indicate the observed $68\%$ and $95\%$ confidence regions estimated in observations for these quantities.
    }
    \label{fig:evolution}
\end{figure}

Figure \ref{fig:evolution} shows examples of how the stellar distribution evolves over time in these simulations.  The velocity dispersion and half-light radius grow steadily over time, depending on the  $\mfdm$ value. For some values, the heating rate is so strong that the evolved dispersion is inconsistent with observed kinematics, regardless of the assumed initial velocity dispersion.

\section{Analysis}
\label{sec:analysis}

Using this suite of simulations, we constrain the FDM mass using simulation-based Bayesian inference.  For each simulation, we first measure the line-of-sight velocity distribution of the test particles as a function of projected radius, $p_{\rm sim}(v_r|r)$ by using 500 random projections of a given simulation snapshot and approximating the resulting mean distribution by a 2D spline. We then use the likelihood to observe velocities $\{v_i\}$ for stars at projected radii $\{r_i\}$
\begin{equation}
    p_{\rm vel} = \prod_i \int dv_i\, p_{\rm sim}(v_i|r_i)\, p_{{\rm obs},i}(v_i),
\label{eqn:vel}
\end{equation}
where $p_{{\rm obs},i}(v_i)$ is a Gaussian of mean and width given by the observed values reported by Refs.\ \cite{Simon2011,Kirby2013}, and the index $i$ runs over all the observed stars for a given galaxy.  

Note that the conditional PDF of the velocity, $p_{\rm sim}(v|r)=p_{\rm sim}(v,r) / p_{\rm sim}(r)$, is undefined at locations where no test particles are found, which can be an issue for very small initial $r_0$.
Since we do not know the spatial dependence of the selection function of spectroscopic targets, we use only the line-of-sight velocity measurements in constructing the likelihood, i.e.\ we do not also model the likelihood to observe a star at radius $r_i$, which in principle is another prediction of each model.  

\begin{figure*}
\centerline{
    \includegraphics[width=\textwidth]{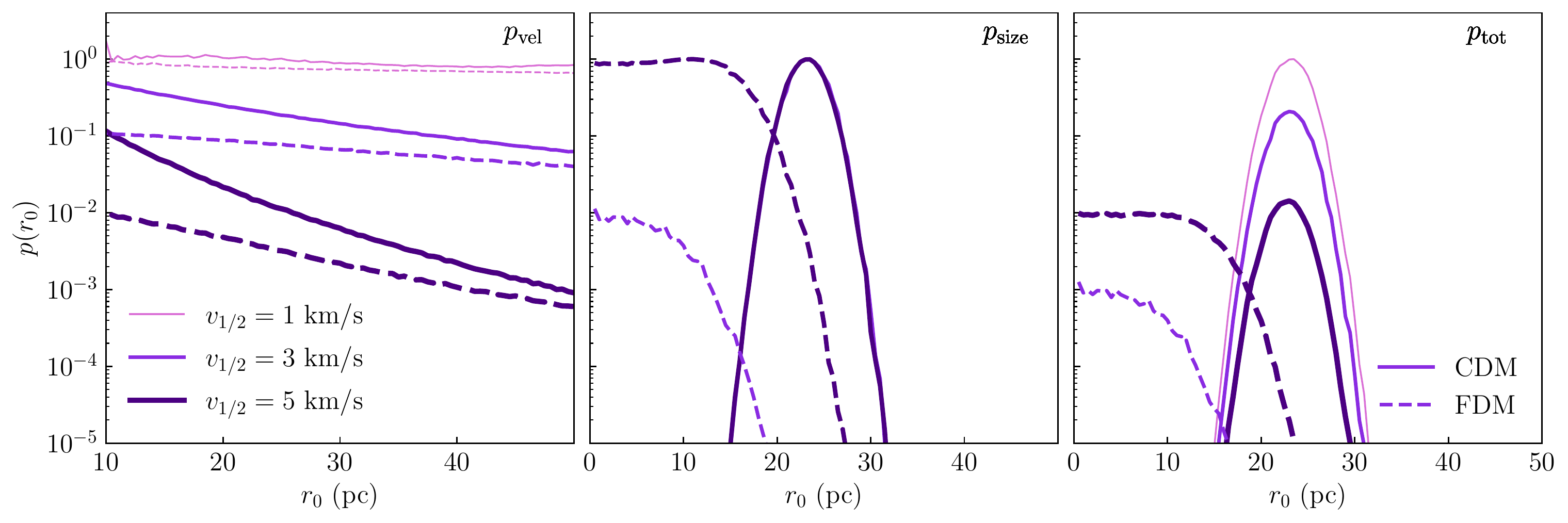}
}
    \caption{Posterior likelihoods as a function of nuisance parameters $r_0$ and \vh, for Segue 2 simulations. From left to right the panels show the likelihoods $p_{\rm vel}$, $p_{\rm size}$, and $p_{\rm tot}$, as defined in Eqns.\ (\ref{eqn:vel}-\ref{eqn:product}). Each panel shows 6 curves, for 3 different values of $\vh$ shown by different colors, and for 2 models: CDM ($\mfdm\to\infty$, solid lines) and FDM with $\mfdm=2\times 10^{-19}$ eV (dashed lines).   Note that for the CDM (smooth potential) models, $p_{\rm size}$ is independent of \vh, while in the middle and right panels, $p_{\rm size}(\vh=1)$ and $p_{\rm tot}(\vh=1)$  have such low values that they are far below the plotted range.}
    \label{fig:rhalf}
\end{figure*}

Instead of modeling the locations of the stars, we evaluate the likelihood for the observed half-light radius to agree with the half-light radius found in each simulation,
\begin{equation}
    p_{\rm size} = \frac{1}{\sqrt{2\pi}\sigma_{1/2}}\,\exp\left[-\frac{(R_{1/2,\rm sim}-R_{1/2,\rm obs})^2}{ 2\sigma_{1/2}^2} \right],
\label{eqn:size}
\end{equation}
where $R_{1/2,\rm obs}$ and $\sigma_{1/2}$ are the observed value and error of the projected 2D half-light radius.  We define the likelihood of each model as the product
\begin{equation}
    p_{\rm tot}({\rm model}) = p_{\rm vel} \times p_{\rm size}.
\label{eqn:product}
\end{equation}
We evaluate $p_{\rm tot}$ for each model, where the model parameters consist of the FDM mass \mfdm, the host halo maximal circular velocity $v_{\rm max}$, the circular velocity at the observed half-light radius \vh, and the initial scale length $r_0$ of the test particles.  Our simulations find that, at fixed \vh, the posterior $p_{\rm tot}$ in Eqn.\ \eqref{eqn:product} is nearly independent of the other NFW parameter $v_{\rm max}$: varying $v_{\rm max}$ over a large range while holding fixed the other parameters produces no discernible difference in $p_{\rm tot}$.  Therefore, for simplicity we set $v_{\rm max}$ to be a fixed multiple of \vh, to avoid marginalizing over an unconstrained, irrelevant parameter.  
We treat $\vh$ and $r_0$ as nuisance parameters and marginalize over them, to give the posterior distribution of \mfdm,
\begin{eqnarray}
    p(\mfdm) &=& \int d\vh\, dr_0\, 
    p_{\rm prior}(\vh,r_0)\nonumber \\
    && ~ \times 
    p_{\rm tot}(\mfdm,\vh,r_0) .
\label{eqn:posterior}
\end{eqnarray}
We use a flat (uniform) prior on $\vh$ and $r_0$ as our fiducial choice, but in practice we find $p_{\rm tot}$ to be so strongly peaked that the specific choice of prior does not significantly affect the resulting constraints, as discussed below (see Table I).  

For the initial stellar scale length $r_0$, we use a uniform prior over the range $0<r_0<37.5$ pc for Segue 1, and $0<r_0<50$ pc for Segue 2.  In both cases, these cover more than the physically plausible range of $r_0$ values, since the upper end already gives an initial $\Rh$ larger than the observed value, even without FDM heating, while the lower end of this range would imply that star formation takes place across a volume even smaller than the size of a single star cluster, which is difficult to reconcile with the observed wide spread in stellar metallicities in these galaxies.  
For \vh, we adopt a uniform prior over the range 1 km/s $<\vh<12.5$ km/s for Segue 1, corresponding to the enclosed mass $10^4 M_\odot < M(\rh) < 1.4 \times 10^6 M_\odot$.   For Segue 2, we assume a uniform prior 1 km/s $< \vh <$ 5 km/s.  For both Segue 1 and Segue 2, the posterior pdf falls off sharply near the upper end of the $\vh$ range (see Fig.~\ref{fig:rhalf}), meaning that we have covered the relevant ranges for the nuisance parameters.

The data that we fit in Eqns.\ \eqref{eqn:vel} and \eqref{eqn:size} are therefore the projected half-light radii, 
$\Rh=24.2 \pm 2.8$ pc and $\Rh=40.5\pm 3.0$ pc for Segue 1 and Segue 2, respectively (see Section~I), 
along with the line-of-sight velocities for individual stars reported by Refs.\ \cite{Simon2011,Kirby2013}. For stars with a single velocity measurement, we adopt the quoted measurement error as the uncertainty.  This neglects the effect of binary motions in the singly-observed stars, which likely leads to an overestimate of the inferred velocity dispersion, thereby making our constraints on \mfdm\ even more conservative.   

For Segue 1, a subset of stars were observed repeatedly, and for those stars we use the weighted average of the measurements. Given that repeated observations of binary stars may reflect velocities in different parts of the orbit, one cannot use standard inverse-variance weighted average that assumes that measurements are drawn from the same parent distribution. To this end, we use the estimators of the weighted mean and its variance proposed by   DerSimonian and Laird \cite{DerSimonian1986}, as given by Eqns.\ (12) and (13) in \cite{Rukhin_2009}. This estimator provides an estimate of the weighted mean and its variance equivalent to the standard one if multiple measurements are consistent with each other, but properly accounts for the variance when they are not.

\section{Results}
\label{sec:results}

As explained in the previous section, we define the posterior likelihood as a product of terms corresponding to stellar kinematics (Eqn.\ \ref{eqn:vel}) and the radial distribution of stars (Eqn.\ \ref{eqn:size}).  In Fig.\ \ref{fig:rhalf}, we show examples of how these terms behave as a function of the nuisance parameters $r_0$ and \vh.  First let us consider the results for simulations in smooth potentials without FDM heating, corresponding to $\mfdm\to\infty$. In the absence of FDM heating, the initial stellar distribution is essentially unchanged in the evolved distribution, except for small initial transients that arise because the locally Maxwellian velocity distribution is not exactly an equilibrium distribution, even when the Jeans equations are satisfied.  Therefore, the evolved $\Rh$ simply scales linearly with the initial $r_0$, and $p_{\rm size}(r_0)$ is a Gaussian function of $r_0$, peaking where $r_0 \approx R_{1/2,{\rm obs}}/1.68$ (recall that $\Rh$ is the half-light radius of the projected 2D surface density, which is different than the 3D half-light radius $\rh$ by about a factor of 1.34).  Note that $p_{\rm size}$ is independent of the assumed halo profile for smooth potentials with no FDM heating, i.e.\ it depends only on $r_0$ and is independent of \vh. In contrast, $p_{\rm vel}$ does depend on the halo profile (parameterized by \vh), and only weakly depends on $r_0$. The product, $p_{\rm tot}=p_{\rm vel}\times p_{\rm size}$, therefore is peaked in both $r_0$ and \vh, and recovers the standard result that $M_{1/2}=3 \sigma_v^2 \rh/G$ \cite{wolf_etal10}, or equivalently in terms of our parameterization,  $\vh = \sqrt{3}\sigma_v$.  For smooth potentials without FDM heating, our analysis therefore boils down to a cumbersome approach of the standard Jeans analysis.

For simulations with a finite \mfdm, FDM heating allows the final evolved $\Rh$ to be different from the initial $\Rh\approx 1.68\, r_0$, giving somewhat different behavior from what is found in smooth potentials.  For very light masses, FDM heating is so strong that stars diffuse out to large distances, causing the final $\Rh$ to exceed the observed value, for any initial $r_0$.  This depends somewhat on \vh, since the local de Broglie wavelength itself depends on \vh.
Namely, the larger values of $\vh$ can reduce FDM heating for a given value of \mfdm.  This is illustrated by the middle panel of Fig.\ \ref{fig:rhalf}, which shows that $p_{\rm size}$ increases as $\vh$ increases.  However, as we saw with the smooth potentials, not all $\vh$ are consistent with the kinematic data, i.e.\ we cannot make $\vh$ arbitrarily large without eventually decreasing $p_{\rm vel}$.  This is illustrated by the left panel, which shows that $p_{\rm vel}$ decreases as $\vh$ increases. 

The product $p_{\rm tot}$ therefore has a peak, and when we marginalize over $r_0$ and $\vh$, we can see that the marginalized posterior pdf will be significantly smaller for $\mfdm=2\times 10^{-19}$ eV than it is for $\mfdm\to\infty$.  
This illustrates how the combination of structural and kinematic data allow us to constrain the FDM mass: the kinematic data basically constrains the potential depth, and the measurement of $\Rh$ then constrains the allowed range of \mfdm.  

\begin{figure}
    \centering
    \includegraphics[width=0.48\textwidth]{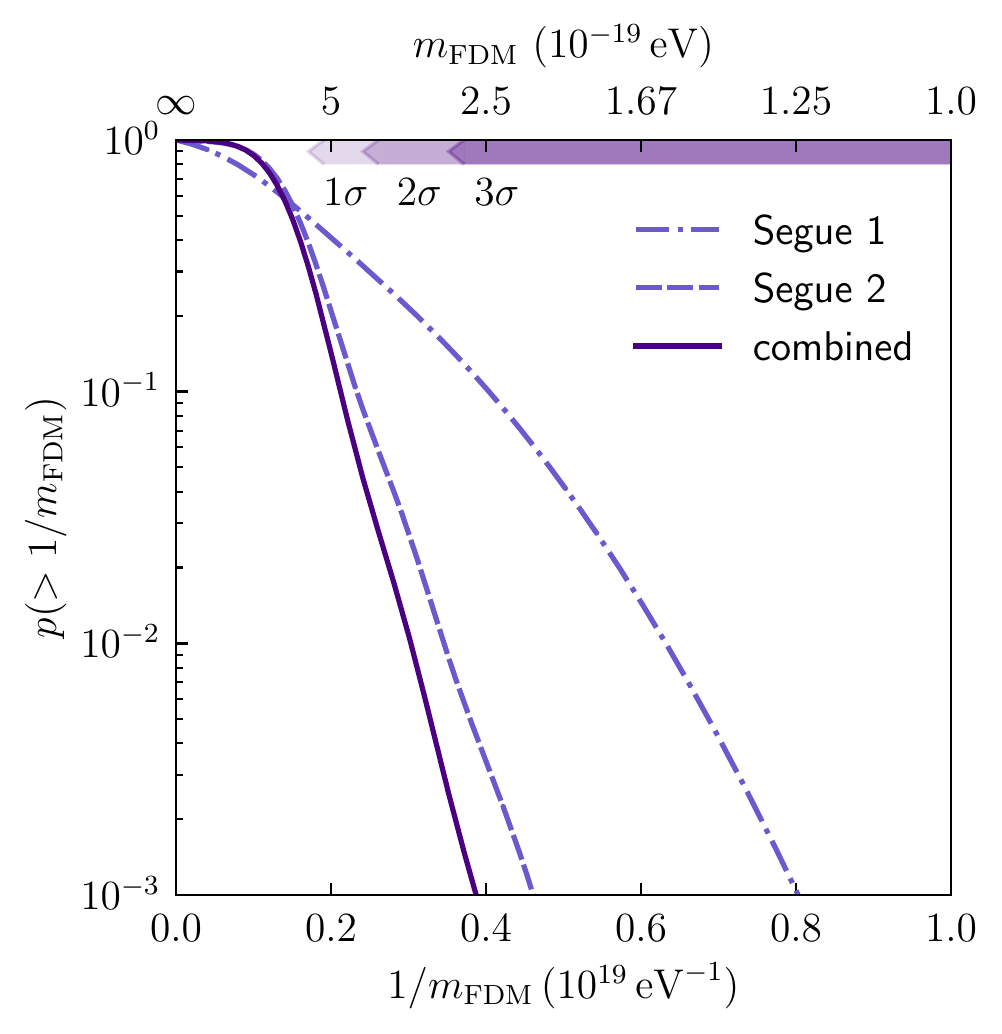}
    \caption{Marginalized posterior likelihood of \mfdm.  Each curve shows the cumulative posterior pdf of $\mfdm^{-1}$, while the arrowheads at top indicate the derived 1, 2, and $3-\sigma$ exclusion regions for joint constraints combining Segue 1 and Segue 2.}
    \label{fig:constraints}
\end{figure}

In Figure \ref{fig:constraints} we show the cumulative probability distribution $p(>1/\mfdm)$ obtained after marginalization over $r_0$ and $\vh$.  As expected from the estimate given in Eqn.\ \eqref{eqn:ballpark}, these galaxies appear to be inconsistent with FDM masses of order $10^{-19}$ eV.  
We show individual constraints  for Segue 1 and Segue 2, and because the observations of these galaxies are independent with uncorrelated uncertainties, we also show the combined posterior pdf shown by the solid line.  Since CDM corresponds to $\mfdm\to\infty$, we show constraints as a function of $\mfdm^{-1}$, assuming a uniform prior on $\mfdm^{-1}$.  Note that this corresponds to a prior $\propto \mfdm^{-2}$, i.e.\ we adopt a prior that heavily favours light FDM masses and strongly disfavours CDM.  Once again, this makes our constraints conservative. 

The posterior pdf gives $\mfdm> 3\times 10^{-19}$ eV at 99\% confidence for the fiducial choice of the prior pdf for $\vh$. Results are not sensitive to the choice of the prior pdf, as shown in Table~\ref{tab:constraints}.   

\begin{table}[b]
\caption{\label{tab:constraints}%
Constraints on the FDM particle mass (in units of $10^{-19}$ eV) for different choices of the $p(\vh)$ prior pdf.
}
\begin{ruledtabular}
\begin{tabular}{cccc}
\textrm{\mfdm\ limit}& 
\textrm{$\rm const$}&
\textrm{$\propto \vh$}&
\textrm{$\propto 1/\vh$}
\\
\colrule
1$\sigma$ & 5.87 & 6.52 & 6.16\\
2$\sigma$ & 4.15 & 4.54 & 4.57\\
3$\sigma$ & 2.86 & 3.22 & 3.35\\
\end{tabular}
\end{ruledtabular}
\end{table}

\section{Discussion}
\label{sec:discussion}

In this paper, we have presented constraints on the mass of ultra-light dark matter particles using kinematic and structural observations of the nearby ultra-faint dwarf galaxies Segue 1 and Segue 2.  Our analysis has made a number of simplifying assumptions that act to make our constraint more conservative, including the neglect of heating effects from a central soliton, the neglect of any initial velocity anisotropy, and the neglect of any binary correction to the kinematic measurements of stars observed only once.  In addition, we have also adopted a prior $p\propto \mfdm^{-2}$ that favors smaller \mfdm\ values.  Nevertheless, we show that stars of Segue 1 and Segue 2 impose a stringent lower limit of $\mfdm> 3\times 10^{-19}$ eV at 99\% confidence.  Below, we discuss implications of this result, as well as potential caveats.

\noindent $\bullet$ \textbf{Galaxies or star clusters?} 
To derive this constraint, we have assumed that these galaxies reside in dark matter halos, i.e.\ we have assumed Segue 1 and 2 are both galaxies, and not star clusters.  For Segue 1, this assumption appears clearly justified, since the inferred dynamical mass vastly exceeds the observed stellar mass.  For Segue 2, however, this assumption may be questioned, since the observed kinematics in Segue 2 are consistent with $\sigma_v=0$.  Indeed, the low dispersion in Segue 2 is the reason why it provides such strong bounds on \mfdm. Although stellar velocities in this galaxy provide only an upper limit on $\vh$, the large spread in metallicity of Segue 2 stars is typical for dark matter dominated dwarf galaxies \cite{Kirby2013}, and is not observed in star clusters.  
Thus, we adopt an assumption that Segue 2 is a dwarf galaxy embedded in a dark matter halo.  

\noindent $\bullet$ \textbf{Tidal stripping:} 
Dynamical heating of stars due to wave interference fringes can be suppressed if a galaxy's DM halo is tidally stripped down to the soliton size. This is because wave interference requires multiple states to be occupied, so when no excited states are present in the wavefunction then FDM heating will be absent \cite{Schive2020}. 

We do not think this is a concern for Segue 1 and Segue 2.
At their current locations, the tidal radii of both satellites are far larger than the observed galaxy sizes \cite{Simon2019}.  Gaia observations of their proper motions indicate that both galaxies are near the pericenters of their orbits \cite{Simon2018}, where tides are strongest.  Therefore, at their present locations, the tidal radii should be close to the smallest they have ever been, and the current tidal radii are far larger than each galaxy's size.   Additionally, the relatively high metallicity of Segue 2, which was viewed as a possible indication of significant tidal stripping by \cite{Kirby2013}, turned out to be typical for UFDs of similar luminosity \cite{Simon2019}.  Similar arguments hold true for Segue 1 as well \cite{Fritz2018}.

At the same time, the soliton size in Segue 1 and 2 cannot be much larger than the observed $r_{1/2}$ of the stars. This is because the size of a DM-dominated stellar system located within a soliton in an isolated halo quickly grows to become comparable to the soliton size after $\lesssim 10$ dynamical times, due to strong dynamical heating by soliton fluctuations \citep{Schive2020, Dutta2021}.  In Segue 1 and Segue 2, the dynamical time is short enough that this process should occur in $\sim 10^8$ yrs, as long as wave interference is present.

Although these galaxies are currently satellites of the Milky Way and their DM halos could be tidally stripped, cosmological simulations of Milky Way sized halos show that nearly all such satellites have been accreted within the past 8 Gyrs (e.g., \cite{dsouza_bell21}). Note also that it would take at least an orbital period ($\approx 1\,$Gyr for Segue 1 and 2 \cite{Simon2018}) to significantly strip their parent halos.  Given that stars in these galaxies were born $\gtrsim 10$ Gyrs ago, they were present when parent halo did not suffer significant tidal stripping for at least $\sim 2-3$ Gyrs, and their sizes should therefore have been inflated to the soliton size during this time. 

These considerations indicate that soliton size can be at most comparable to the current extent of the galaxy's stellar system. Given that the tidal radii of Segue 1 and Segue 2 are estimated to be much larger than this extent, then wave interference fringes and soliton oscillations should be present in the halos of these galaxies, validating the constraints derived in this paper.

\noindent $\bullet$ \textbf{Other galaxies:} 
Although we have only shown results for two of the smallest UFD galaxies, additional constraints at lighter masses are provided by other galaxies (including other UFDs), which span a wide range of sizes and velocities.  FDM masses below $10^{-21}$ eV are already excluded by the mere existence of halos hosting the smallest Milky Way satellites \cite{Nadler2021}, as well as the small-scale Lyman-$\alpha$ forest power spectrum \cite{Hui2021}.  Using the reasoning behind Eqn.\ \eqref{eqn:ballpark}, we can see that ultra-thin galaxies like DDO 168  with circular velocities in the range 60-100 km/s, sizes of few kpc, and vertical velocity dispersions of $\sim 10$ km/s \cite{ddo168} can exclude FDM masses of this same order, $m\gtrsim 10^{-21}\,$eV.  Relatively large UFDs like Carina, with velocity dispersions $\sim 7$ km/s and sizes $\sim 300\,$pc can extend the bound to $m\gtrsim 5\times 10^{-21}\,$eV while somewhat smaller dwarfs like Bo\"otes I with $\Rh\approx 200$ pc and $\sigma_v\approx 5$ km/s should impose $m>10^{-20}$ eV.  Even smaller UFDs like Leo IV ($\Rh\approx 100\,$pc, $\sigma_v\approx 3\,$km/s) can push this further to $m>3\times 10^{-20}$ eV.  Obviously, these bounds are not competitive with the constraints imposed by the smallest galaxies, but we can see that even if we removed the constraints from Segue 1 and 2, FDM with $m<10^{-20}\,$eV would remain excluded by these other UFDs.

\noindent $\bullet$ \textbf{Soliton effects:} 
As mentioned above, our analysis neglects soliton effects. 
It is worthwhile to consider how our constraints would change if we included solitons in our simulations.  As noted in \S\ref{sec:sims}, it is straightforward to add solitons to our simulated halos, simply by adding them to the input density profile. This method gives fluctuations in the soliton density and centroid location similar in size to what is found in self-consistent Schr\"odinger-Poisson simulations \cite{Yavetz2021, Li2021}, and therefore should provide an estimate of the effect of the soliton heating.  

We have not simulated the entire parameter range covered by our non-soliton simulations, but the results of a limited number of soliton simulations that we have performed indicate that solitons significantly enhance FDM heating in the central regions of halos, as expected from previous work \cite{Dutta2021}.  In Segue 1 and Segue 2, this enhanced heating acts to further worsen the fit to the measured data.  We find that the posterior likelihood for $\mfdm=10^{-19}$ eV is reduced by an order of magnitude, compared to our no-soliton simulations, when solitons of mass and size given by the scaling relation of Ref.\ \cite{Schive:2014hza} are included.

\noindent $\bullet$ \textbf{Other UFD analyses:} 
This result appears to be inconsistent with a recent analysis of 18 UFDs (including Segue 1 and Segue 2) which claims to exclude CDM in each of those 18 galaxies \cite{Hayashi2021}.  The basis for that claim is a Jeans analysis that apparently finds significant evidence for a massive soliton in each galaxy, which is then translated into a measurement of $\mfdm$ using the soliton core-halo scaling relation (see their Fig.\ 1).  One significant issue is that their analysis is not self-consistent, due to neglect of the FDM heating effects that we discuss in this paper.  As noted above, inclusion of soliton effects (meaning both the effect on the average gravitational potential and also the heating effect) significantly {\em worsens} the fit to observations of Segue 1 and 2 for $\mfdm \sim 10^{-19}$ eV.  It is possible that neglect of heating might explain why Ref.\ \cite{Hayashi2021} reach such different conclusions analyzing the same data used here.  We do note, however, that it is difficult to understand how their analysis can significantly detect solitons at $>2\sigma$ in each of the 18 galaxies, while also remaining uncertain in the mass and size of those detected solitons by many orders of magnitude.  A more recent Jeans analysis of 5 UFDs \cite{Zoutendijk2021b} (including some also present in the sample of \cite{Hayashi2021}) finds no evidence of solitons in any of the 5, in contradiction to the results of Ref.\ \cite{Hayashi2021}.  For these reasons, we do not currently believe the inconsistency with Ref.\ \cite{Hayashi2021} is a reason to doubt our derived lower limit on $\mfdm$.

\noindent $\bullet$ \textbf{Is FDM ruled out?} 
As noted above, our constraints exclude the mass range formally defined as fuzzy dark matter.
In a more practical sense, the constraint derived here effectively excludes the ``fuzzy'' regime of ultra-light dark matter.  Using the linear transfer function given in Ref.\ \cite{Hu:2000ke}, the linear power spectrum in ultra-light DM models satisfying this constraint on $\mfdm$ will be identical to the $\Lambda$CDM linear power spectrum out to $k \sim 200\,h\,$Mpc$^{-1}$.\footnote{Note, however, that with sufficient fine-tuning, the large misalignment mechanism \cite{Arvanitaki2020} can modify this conclusion.} This means that  ultra-light dark matter will be indistinguishable from CDM in the Lyman-$\alpha$ forest, which can probe the linear matter power spectrum to $k\lesssim 10$ Mpc$^{-1}$.  Similarly, the FDM correction to the halo mass function \cite{Schive16a} for this particle mass is small for $M\gtrsim 2\times 10^5\, h^{-1} M_\odot$, which means that FDM models are constrained to be indistinguishable from the CDM using probes of halo substructure like strong lensing or tidal streams \cite{Hezaveh2016,Erkal2016}.  To our knowledge, there are no existing galactic observations that can distinguish ultra-light dark matter from CDM with the limits imposed by Segue 1 and Segue 2 observations.

\noindent $\bullet$ \textbf{Future directions:} 
This bound on the FDM mass can be potentially improved.  For the diffusive heating effect studied here, observing additional galaxies would be unlikely to provide significant improvements, unless those galaxies have sizes or velocity dispersions significantly smaller than Segue 1 and 2 (see Eqn.\ \ref{eqn:ballpark}).  A more promising way to strengthen these constraints would be to account for the effects of central solitons, which were ignored in our analysis due to the uncertainty in the scaling between soliton mass and halo mass.  

As noted above, our simulations indicate that solitons of mass given by the scaling relation of Ref.\ \cite{Schive:2014hza} can significantly enhance FDM heating effects.  These simulations also point to a distinctive signature of soliton heating, namely a sharp increase in stellar velocity dispersion in the vicinity of the soliton.  This suggests that future observations could significantly strengthen bounds on ultra-light DM by measuring kinematics of stars at the very centres of UFDs, if future theoretical work does firmly establish that all FDM halos should have solitons sufficiently massive to enhance stellar heating.

Besides galaxies, observations of supermassive black holes could further strengthen the bound on  ultra-light bosonic particles.  The FDM mass range probed by the UFDs is similar in magnitude to the mass range that can generate superradiance in the largest known black holes \cite{Arvanitaki2010,Arvanitaki2011,Du2022}.  Detection of large spins in these black holes, as well as in lower mass black holes, could be used to further narrow the range of allowed dark matter particle masses by many orders of magnitude beyond what we derived here.

\acknowledgements
We thank Vasily Belokurov, Dhruba Dutta Chowdhury, Lam Hui, Xinyu Li, Simon May, Luna Zagorac, and Bas Zoutendijk for helpful discussions.
ND is supported by the Centre for the Universe at Perimeter Institute, and by MEXT KAKENHI Grant Number JP20H05861.
Research at Perimeter Institute is supported in part by the Government of Canada through the Department of Innovation, Science and Economic Development Canada and by the Province of Ontario through the Ministry of Colleges and Universities. This research was enabled in part by resources provided by Compute Ontario and Compute Canada and computational resources provided by the University of Chicago Research Computing Center. 
This work has made use of the {\tt GSL} \cite{GSL}, {\tt FFTW} \cite{FFTW}, {\tt SHTns} \cite{shtns}, {\tt NumPy} \citep{numpy_ndarray}, {\tt SciPy} \citep{scipy}, and {\tt Matplotlib} \citep{matplotlib} libraries, and we thank the respective authors for making their software publicly available. 
We have also used the Astrophysics Data Service (\href{http://adsabs.harvard.edu/abstract_service.html}{\tt ADS}) and \href{https://arxiv.org}{\tt arXiv} preprint repository extensively during this project and the writing of the paper.

\newcommand{\jcap}{J.\ Cosm.\ Astroparticle Phys.}
\newcommand{\mnras}{Mon.\ Not.\ Roy.\ Astron.\ Soc.}
\newcommand{\aj}{Astron.\ J.}
\newcommand{\apjl}{Astrophys.\ J.\ Lett.}
\newcommand{\aap}{Astron.\ Astrophys.}
\newcommand{\araa}{Ann.\ Rev.\ Astron.\ Astrophys.}

\bibliography{fdm}

\end{document}